\documentclass{iopart}
\usepackage{amsfonts,amssymb} 
\usepackage[fleqn]{amsmath}
\usepackage{graphicx,color}
\usepackage{lineno}

\def\be{\begin{equation}}
\def\ee{\end{equation}}
\def\bdm{\begin{displaymath}}
\def\edm{\end{displaymath}}

\renewcommand{\bf}{ }
\ioptwocol

\begin{document}
\title{Regularized $\kappa$-distributions with non-diverging moments}
\author{K. Scherer$^{1,2}$, H. Fichtner$^{1,2}$, M. Lazar$^{1,3}$}
\address{$^{1}$Institut f\"ur Theoretische Physik, Lehrstuhl IV: Weltraum-
  und Astrophysik, Ruhr-Universit\"at Bochum, D-44780 Bochum, Germany}
\address{$^{2}$Research Department, Plasmas with Complex Interactions,
  Ruhr-Universit\"at Bochum, 44780 Bochum, Germany}
\address{$^{3}$Centre for Mathematical Plasma Astrophysics, Celestijnenlaan 200B, 3001 Leuven Belgium }
\ead{kls@tp4.rub.de}

\date{\today}

\begin{abstract}
  For various plasma applications the so-called (non-relativistic) $\kappa$-distribution is 
	widely used to reproduce and interpret the suprathermal particle populations exhibiting a 
	power-law distribution in velocity or energy. Despite its reputation the standard 
	$\kappa$-distribution as a concept 
	is still disputable, mainly due to the velocity moments $M_{l}$ which make possible
	a macroscopic characterization, but whose existence is restricted only to low orders $l 
	< 2\kappa-1$. In fact, the definition of the $\kappa$-distribution itself is conditioned by the
	existence of the moment of order $l=2$ (i.e., kinetic temperature) satisfied only for $\kappa > 3/2$.  
	In order to resolve these critical limitations we introduce the regularized $\kappa$-distribution
	with non-diverging moments. For the evaluation of all velocity moments a general analytical 
	expression is provided enabling a significant step towards a macroscopic (fluid-like) 
	description of space plasmas, and, in general, any system of $\kappa$-distributed particles.
\end{abstract}


\submitto{EPL}
\maketitle
\section{\label{sec:level1}Introduction}
%%%%%%%%%%%%%%%%%%%%%%%%%%%%%%%%%%%%%%%%%%

The $\kappa$-distribution was introduced into space physics by Olbert \cite{Olbert-1968}, 
see also Vasyliunas \cite{Vasyliunas-1968}, in order to describe spectral measurements of 
magnetospheric energetic electrons. Since then it has not only been employed successfully 
for quantitative and fundamental interpretations in space plasmas \cite{Treumann-Jaroschek-2008, 
Pierrard-Lazar-2010, Beck2017}, but also applied in other fields of plasma physics 
\cite{Webb-etal-2012, Elkamash-Kourakis-2016}. Furthermore, its relation to 
Student-t probability function \cite{Gosset-1908} and to Pearson's 
type IV distribution \cite{Pearson-1916} has been reported
\cite{Abdul-Mace-2014}, its seeming ambiguity regarding an interpretation in terms of thermal 
speed and temperature has been discussed \cite{Lazar-etal-2016, Ziebell-Gaelzer-2017}, and the 
possibility of different Maxwellian limits have been pointed out \cite{Lazar-etal-2017}. 

While the above studies have contributed to put the $\kappa$-distribution,
which originally had been introduced in an {\it ad hoc} manner, on
solid physical grounds, they did not remove one major unphysical
feature, namely the occurrence of diverging velocity moments. The pure
power-law character of the standard $\kappa$-distribution
unavoidably implies that only a finite number of velocity moments
exists. This severe limitation restricts the derivation of a closed
system of fluid equations and, thus, prevents the $\kappa$-distribution
from being a contradiction-free representation of a phase space
distribution function of a physical system with a high-velocity (suprathermal)
component relative to a Maxwellian plasma.

In order to remove this deficiency, we introduce the `regularized' $\kappa$-distribution, which 
fulfils the same purpose as the standard one but, additionally, has the property that {\it all} 
velocity moments remain finite. Beyond the fact that all of these moments can be computed 
analytically as a consequence of the new distribution's exponential cut-off, we show that such 
a cut-off is not purely ad-hoc but appears to be a common and natural change of a power-law 
behaviour that cannot extend to arbitrarily high velocities. The improved distribution 
resolves further apparently unphysical limitations regarding the range of allowed $\kappa$-values. 
Finally, we demonstrate that results obtained previously by using the standard 
$\kappa$-distribution remain unchanged if they were derived within the framework of kinetic 
theory. Changes have to be expected, however, as soon as moments of the distribution are used.

%%%%%%%%%%%%%%%%%%%%%%%%%%%%%%%%%%%%%%%%%%%%%%%%%%%%%%%%%%%%%%%%%%%
\section{\label{sec:level2}{\large $\kappa$}- and Maxwell-distributions}
%%%%%%%%%%%%%%%%%%%%%%%%%%%%%%%%%%%%%%%%%%%%%%%%%%%%%%%%%%%%%%%%%%%

The standard (isotropic, steady-state) $\kappa$-distribution (SKD) of plasma particles of speed $v$,
with number density $n = n(\vec{r})$ depending on location $\vec{r}$, is defined as
\begin{eqnarray}
  f_{\kappa}(\Theta,v) = n N_{\kappa} \left(1+ \frac{v^{2}}{\kappa \Theta^{2}}\right)^{-\kappa-1}   
	\label{eq:k}
\end{eqnarray}
The constant $N_{\kappa}= {\Gamma(\kappa)}/{\Gamma\left(\kappa-\frac{1}{2}\right)}/(\sqrt{\pi\kappa}\Theta)^{3}$ 
normalizes the distribution such that $\int f_{\kappa}(\Theta, v)\mathrm{d}^{3}v = n$, 
and $\Theta$ is a `thermal speed' normalizing the particle speed. Via the limit $\kappa\to\infty$ this
family of functions can be related to a Maxwellian distribution (MD), in the generic form
\begin{eqnarray}
  \label{eq:M}
  f_{M}(w,v) = n N_{M} \exp\left\{-\frac{v^{2}}{w^{2}}\right\}
\end{eqnarray}
with $n$ as above and the thermal speed $w$ normalizing the particle speed, 
so that $N_{M}= {1}/({\sqrt{\pi} w})^{3}$ normalizes the distribution to $n$ again. If $\Theta$ is 
independent of $\kappa$, the Maxwellian limit with $w = \Theta$ is obtained, and it approaches the 
low-energy core of the $\kappa$-distribution \cite{Lazar-etal-2015}. The characterization of 
suprathermal populations and their 
effects becomes then straightforward by contrasting Eqs.~\ref{eq:k} and \ref{eq:M} \cite{Lazar-etal-2016}. 
We name this limit the relevant Maxwellian distribution (RMD).

Alternatively to the SKD, $\Theta$ can be 
a function of $\kappa$ by $\Theta = (2 {\rm k_B} T_\kappa /m)^{0.5} (1-1.5/\kappa)^{0.5}$ resulting
from the definition of the kinetic temperature as the second order moment $T_\kappa = (m/2 {\rm k_B}) 
\int f_{\kappa} v^2 \mathrm{d}^{3}v$, which needs to be independent of $\kappa$ ($m$ is the 
mass of plasma particle and ${\rm k_B}$ is the Boltzmann constant). In this case, 
$w = (2 {\rm k_B} T_\kappa /m)^{0.5}$, but the relevance of a Maxwellian limit with $T_M = T_\kappa$ 
is not clear yet \cite{Lazar-etal-2016}. 

While, due to the exponential cut-off of the MD, all its (isotropic) velocity moments 
$4\pi \int f_M(w,v) v^{l+2} dv$ (with the integer $l$ defining the $l$-th moment) remain 
finite, this is not true for the SKD, for which $4\pi \int f_\kappa(\Theta,v) v^{l+2} dv < \infty$ 
only if $l<2\kappa+1$. Evidently, in order to avoid diverging moments, a power-law distribution 
needs to be modified such that it exhibits an exponential cut-off towards higher speeds.

\begin{figure}[t!]
  \includegraphics[width=.50\textwidth]{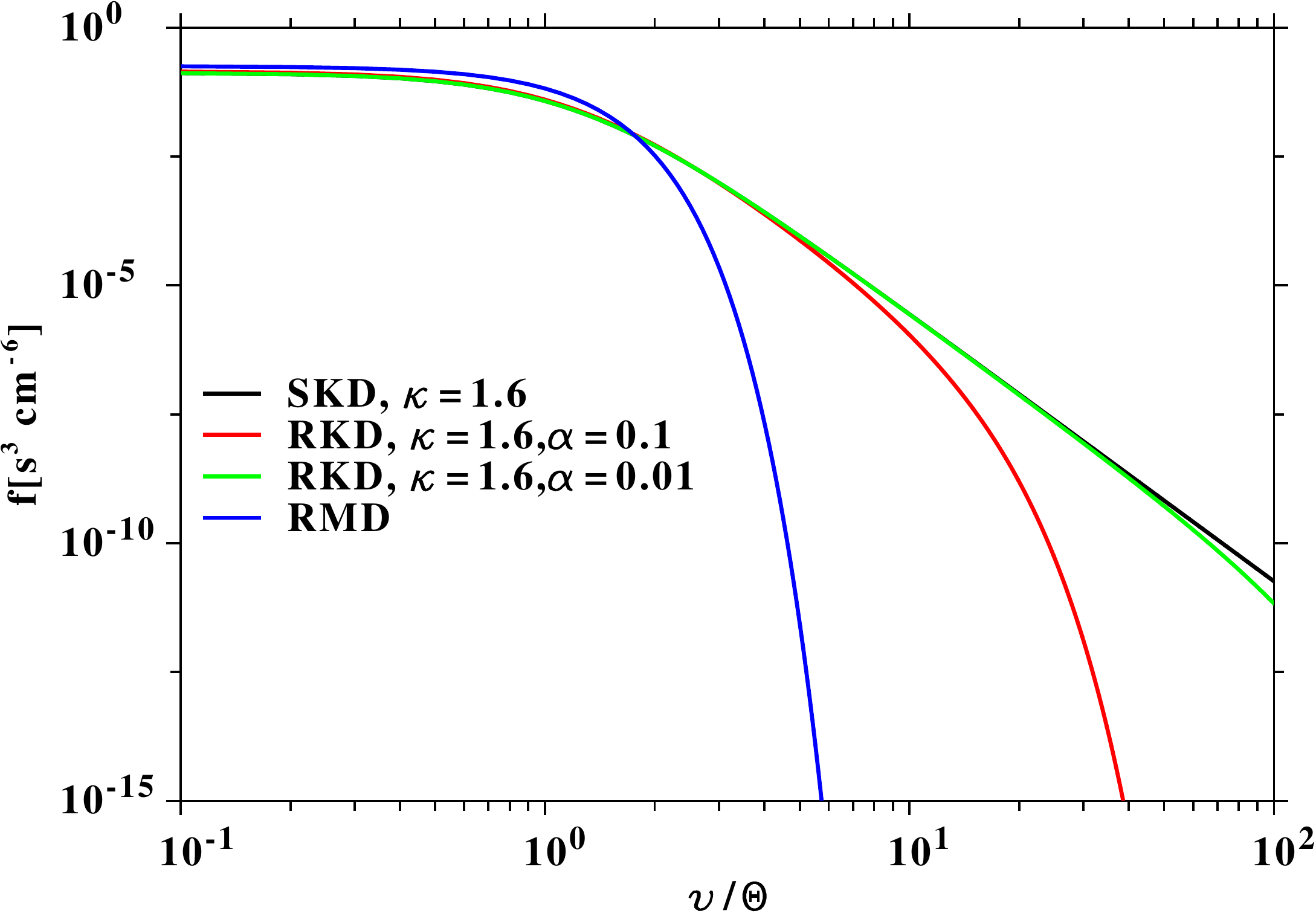}
  \caption{Comparison of the SKD (black line, for $n = 1$\,cm$^{-3}$, $\kappa=1.6$), 
	RKD for $\alpha=0.1$ (red line) and $\alpha=0.05$ (green line), and the RMD for 
	$\kappa \to \infty$ and $\alpha = 1$ (blue line). While the abscissa corresponds 
	to the normalized speed $x=v/\Theta$, the ordinate is in units of particles per phase 
	space volume.} \label{fig:1}
\end{figure}
%

%%%%%%%%%%%%%%%%%%%%%%%%%%%%%%%%%%%%%%%%%%%%%%%%%%%%%
\section{The regularized {\large$ \kappa$}-distribution}
%%%%%%%%%%%%%%%%%%%%%%%%%%%%%%%%%%%%%%%%%%%%%%%%%%%%%

Guided by the properties of the MD we introduce the regularized $\kappa$-distribution (RKD):
\begin{eqnarray}\label{eq:mod}
  f(\kappa,\alpha,\Theta,v) &\equiv& \frac{N_{r\kappa}}{n N_{M}N_{\kappa}}
	f_{\kappa}(\Theta,v) f_{M}(\alpha,\Theta,v) \\\nonumber
  &&\hspace*{-1.5cm} = n N_{r\kappa} \left(1 + \frac{v^{2}}{\kappa \Theta^{2}}\right)^{-\kappa-1}
                       \exp\left\{-\frac{\alpha^{2} v^{2}}{\Theta^{2}}\right\}\\\nonumber
  &&\hspace*{-1.5cm} = n N_{r\kappa} \ _{1}F_{0}\left([\kappa+1],[];-\frac{v^{2}}
                      {\kappa \Theta^{2}}\right)
                       \exp\left\{-\frac{\alpha^{2}{v}^{2}}{{\Theta}^{2}}\right\}
\end{eqnarray}
where $f_{M}(w,v)=f_{M}(\alpha,\Theta,v)$ is provided by Eq.~\ref{eq:M} for 
$w = \Theta/\alpha$, $_{1}F_{}$ is a hypergeometric function {\bf (see
appendix)} and the normalisation constant 
$N_{r\kappa} = N_{r\kappa}(\alpha,\Theta)$ is given explicitly below. 
Obviously, the RKD has a power-law behaviour at low and intermediate 
speeds, and an exponential cut-off at higher speeds, regulated by the cut-off parameter 
$\alpha$.

%%%%%%%%%%%%%%%%%%%%%%%%%%%%%%%%%%%%%%%%%%%%%%%%%%%%%%%%%%%%%%%
\subsection{Limits of the regularized $\kappa$-distribution}
%%%%%%%%%%%%%%%%%%%%%%%%%%%%%%%%%%%%%%%%%%%%%%%%%%%%%%%%%%%%%%%

It is convenient to introduce a dimensionless argument $x=v/\Theta$ so that 
\begin{eqnarray}\label{eq:mod2}
  f(\kappa,\alpha,\Theta,x) &=& n N_{r\kappa}
                                \left(1 + {x^{2}}/{\kappa}\right)^{-\kappa-1}\exp\left\{-\alpha^{2} x^{2}\right\}
                                \nonumber\\
                            &&\hspace*{-1.7cm}= n N_{r\kappa} \ _{1}F_{0}\left([\kappa+1],[];-{x^{2}}/{\kappa}\right)
                                \exp\left\{-{\alpha^{2}{x}^{2}}\right\}
\end{eqnarray}
Fig.~\ref{fig:1} illustrates the differences between the RMD, SKD and the RKD 
(for two values of the cut-off parameter, $\alpha = 0.1$, and 0.05. For low (normalised) speeds, 
i.e.\ $x<1/\alpha$, the deviations from the SKD are small, while for $x>1/\alpha$ the exponential 
function cuts off the SKD. In the limit of $x\gg \kappa$ the RKD can be 
approximated by 
\begin{eqnarray}
  \label{eq:pl1}
  f(\kappa,\alpha,\Theta, x) \approx  n N_{r\kappa} \kappa^{\kappa+1} x^{-2(\kappa+1)} 
	\exp\left\{-\alpha^{2} x^{2}\right\}
\end{eqnarray}
Second, in the $\alpha \to 0$ the SKD is recovered, and, third, in the 
limit $\kappa \rightarrow \infty$ the RKD corresponds to a Maxwellian distribution of the form
\begin{eqnarray}
\lim\limits_{\kappa\rightarrow\infty}f(\kappa,\alpha,\Theta,v) 
= n N_{r\infty} \exp\left\{-(1+\alpha^{2})\frac{v^{2}}{\Theta^{2}}\right\}, 
\end{eqnarray}
which is defined as in Eq.(\ref{eq:M}) with the choice
$w = \Theta/ \sqrt{1+\alpha^{2}}$, and the RMD is obtained for $\alpha \to 0$.

%%%%%%%%%%%%%%%%%%%%%%%%%%%%%%%%%%%%%%%%%%%%%%%%%%%%%%%%%%%%%%%%%%%%
\subsection{Properties of the regularized $\kappa$-distribution}
%%%%%%%%%%%%%%%%%%%%%%%%%%%%%%%%%%%%%%%%%%%%%%%%%%%%%%%%%%%%%%%%%%%%
1. For a given data set exhibiting a power-law behaviour, fits to measurements will yield the same results
  for the SKD and the RKD, if $\alpha$ can be chosen sufficiently small so that the cut-off occurs well beyond the 
  range of observations. This is most often the case because most instruments take measurements within a finite 
  velocity/speed interval. The fact that such results are insensitive to the use of the SKD or the RKD is illustrated
  below at the example of Langmuir waves. 
  
2. The RKD is well-defined for all positive $\kappa$, i.e.\ resolves the 
divergence of the second-order moment of SKD, occurring for $\kappa \leqslant 3/2$. 

3. Most importantly, all velocity moments of the RKD are finite (as shown in the next section), 
allowing to employ the RKD as a valid basis for a complete hydrodynamical description of a 
$\kappa$-distributed plasma.  %with a significant suprathermal particle population.  

%%%%%%%%%%%%%%%%%%%%%%%%%%%%%%%%%%%%%%%%%%%%%%%%%%%%%%%%%%%%%%%%
\subsection{Moments of the regularized $\kappa$-distribution}
%%%%%%%%%%%%%%%%%%%%%%%%%%%%%%%%%%%%%%%%%%%%%%%%%%%%%%%%%%%%%%%%

The isotropic velocity moments $M_{l}$ with $l\in\{0,1,2,...\}$ of the RKD are defined as 
\begin{eqnarray}\label{eq:2}
  M_{l}(\kappa,\alpha,\Theta) &=& \int v^{l}f(\kappa,\alpha,\Theta,v)\mathrm{d}^{3}v \\\nonumber
  &=& 4\pi \int\limits_{0}^{\infty}v^{\nu}f(\kappa,\alpha,\Theta,v)dv \\\nonumber
  &\equiv& n N_{r\kappa} I(\kappa,\alpha,\nu,\Theta)
\end{eqnarray}
where $\nu = l+2$, and $4\pi$ and $v^2$ are additional factors resulting from the 
integration in spherical coordinates.
The analytical solution of the integral $I(\kappa,\alpha,\nu,\Theta)$ and, thus, for all moments is given by
\begin{align}\label{eq:Ikant} 
  I(\kappa,\alpha,\nu,\Theta) =&  4\pi\,\Theta^{1+{\nu}}\Bigg\{ \\ \notag
% c1
  &- C_{1} \frac {\Gamma \left( a \right) \Gamma \left(b_{1} \right) }{4\Gamma \left( \kappa+1\right)}
    \mbox{$_1$F$_1$}(a;\,-b_{0};\,\alpha^{2}\kappa)   \\ \notag
%c2
  &- C_{2} \frac {\Gamma \left( a \right) \Gamma \left(b_{2} \right) }{8\Gamma \left(\kappa+1 \right)}
   \mbox{$_1$F$_1$}(a;\,-b_{1};\,\alpha^{2}\kappa) \\ \notag 
%c3
  &+ C_{3} \frac{1}{2}\Gamma \left(-b_{3}\right)  \mbox{$_1$F$_1$}(\kappa;b_{4};\,\alpha^{2}\kappa) \\ \notag
%c4
  & -C_{4} \frac{1}{4} \Gamma \left(-b_{4} \right) {\mbox{$_1$F$_1$}(\kappa;\,b_{5};\,\alpha^{2}\kappa)}
     \Bigg\} 
\end{align}
with
\begin{eqnarray}\nonumber
  a &=& \frac{\nu - 1}{2},\qquad b_{i} = \frac{2\kappa-\nu +(2 i
  -5)}{2}\\\nonumber
 \mathrm{and}\\\nonumber
  C_{1} &=& \alpha^{2}\left(2\alpha^{2} \kappa+ \nu - 1 \right)\left(2 -
            \kappa\right)\kappa^{\frac{\nu+3}{2}}\\\nonumber
  C_{2} &=& \left(4 \alpha^{4}\kappa^{2}+\left[(4\nu - 4) \alpha^{2} -2
      \nu +2\right]k +\nu^{2}-1
            \right)\kappa^{\frac{\nu+1}{2}}\\\nonumber
  C_{3} &=& \kappa^{\kappa+1}
            \alpha^{1+2\kappa-\nu}(\alpha^{2}+1)\\\nonumber
  C_{4} &=& (\nu-3)\kappa^{\kappa+1}\alpha^{3+2\kappa-\nu}
\end{eqnarray}
Formula (\ref{eq:Ikant}) is the central result of the paper, as it allows to calculate {\it all} 
velocity moments of the RKD associated with a SKD. Further contrasting to the SKD, which 
is applicable only for $\kappa > 3/2$, the RKD is defined for all $\kappa>0$, and all moments remain
finite for any positive $\kappa > 0$ and $\alpha>0$. {\bf
  \mbox{$_1$F$_1$} is the confluent hypergeometric function or Kummer
  function (see appendix).}

The explicit form of the normalization constant $N_{r\kappa}$ can be provided 
by observing that $n = M_0(\kappa,\alpha,\Theta)$, so that 
\begin{eqnarray}   \label{eq:norm_mk}
  N_{r\kappa} = \frac{1}{I(\kappa,\alpha,2,\Theta)}
\end{eqnarray}
which in the case $\alpha=0$ reduces to $N_{\kappa}$, see Eq.(\ref{eq:k}), and for $\kappa \rightarrow 
\infty$ to $N_{M}$, see Eq.(\ref{eq:M}).

%%%%%%%%%%%%%%%%%%%%%%%%%%%%%%%%%%%%%%%%%%%%%%%%%%%%%%%%%%%%%%%%%%%%
\subsection{Velocity normalization and temperature definition}
%%%%%%%%%%%%%%%%%%%%%%%%%%%%%%%%%%%%%%%%%%%%%%%%%%%%%%%%%%%%%%%%%%%%

The two Maxwellian limits discussed in sec.~II suggest two alternatives for interpreting an SKD. 
The first, claiming use of a $\kappa$-independent $\Theta$ (thermal speed), implies a $\kappa$-dependent 
temperature $T_\kappa = T_\kappa (\kappa)$ and admits the RMD limit for $\kappa \to \infty$ \cite{Leubner-2000,
Langmayr-2005, Lazar-etal-2015}. The 
second option is to assume a $\kappa$-independent temperature, which is usually convenient in 
computations. While there might be systems constrained to evolve under conditions of constant temperature
\cite{Yoon-2014}, the more relevant case appears to be the former one \cite{Lazar-etal-2015,Lazar-etal-2016}. 
Also for the RKD the first choice is appropriate since the exponential cut-off should be independent of 
$\kappa$, implying that $\Theta$ cannot depend on $\kappa$. Consequently, the RKD is defined with a 
$\kappa$-dependent temperature, as follows. For a Maxwellian-distributed plasma one can define the temperature by
\begin{eqnarray}
  \label{eq:temp1}
  T_{M} = \frac{m P_{M}}{k_{B} n}
\end{eqnarray}
with $P_{M}$ being the second-order moment of the Maxwellian distribution function, 
which is the pressure (per mass unit) or the energy density. 

We define the temperature of the RKD analogously: 
\begin{eqnarray}
  \label{eq:temp}
T_{r\kappa} &\equiv& T_{r\kappa}(\kappa,\alpha,\Theta)  
                     \equiv \frac{m P_{r\kappa}}{n k_{B}}\\\nonumber 
  &=&m \frac{N_{r\kappa}}{k_{B}} I(\kappa,\alpha,4,\Theta) =\frac{m}{k_{B}}\,
   \frac{I(\kappa,\alpha,4,\Theta)}{I(\kappa,\alpha,2,\Theta)}
\end{eqnarray}
The integral $I(\kappa,\alpha,2,\Theta)$ in the denominator is the same defining the normalization 
factor $N_{r\kappa}$ in Eq.~\ref{eq:norm_mk}, and the RKD pressure $P_{r\kappa}$ reads
\begin{eqnarray}
P_{r\kappa} = n \frac{I(\kappa,\alpha,4,\Theta)}{I(\kappa,\alpha,2,\Theta)}.
\end{eqnarray}

\begin{figure}[t!]
%\hspace*{-0.4cm}
  \includegraphics[width=0.44\textwidth]{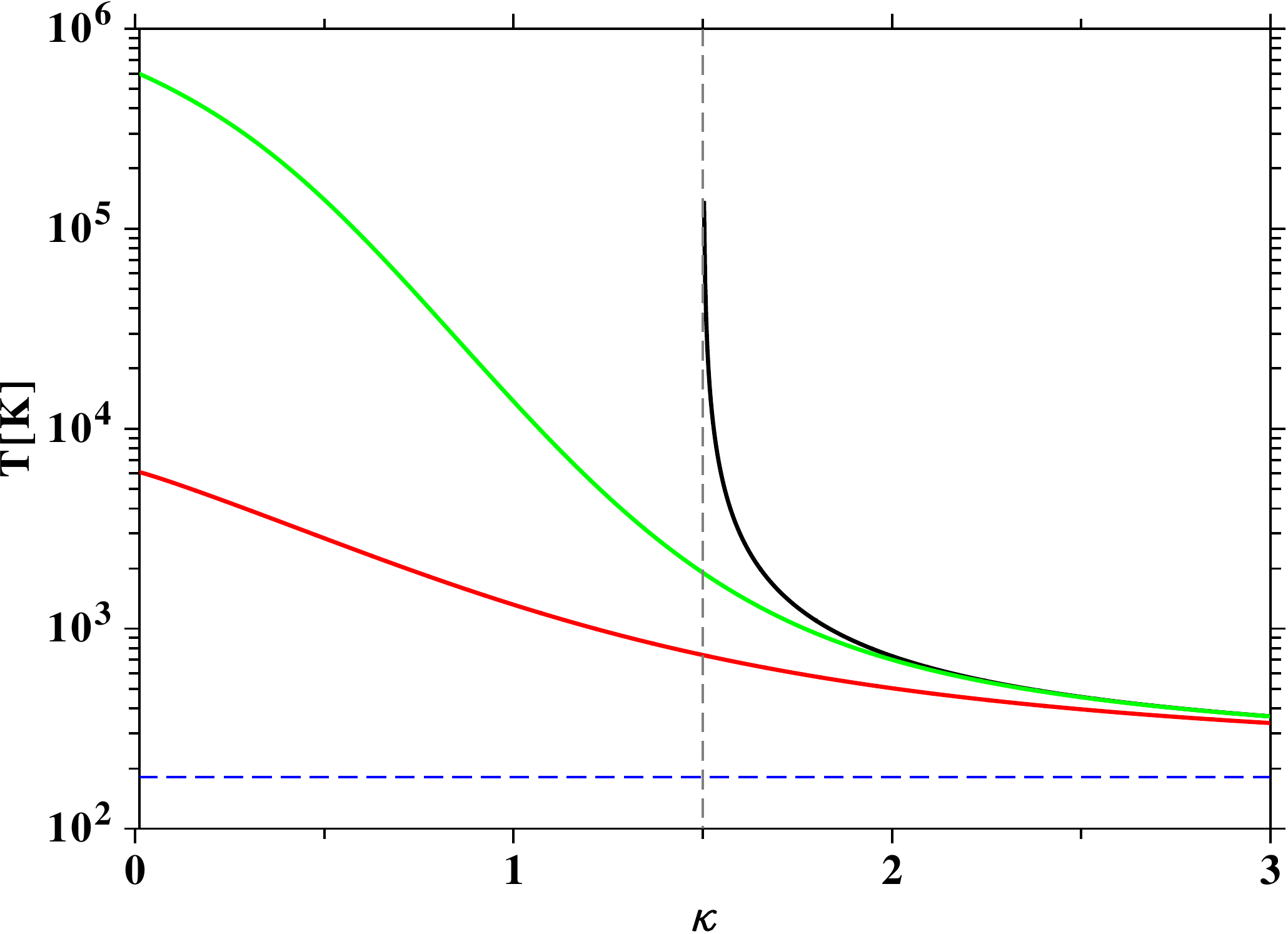}    \\
  \includegraphics[width=0.44\textwidth]{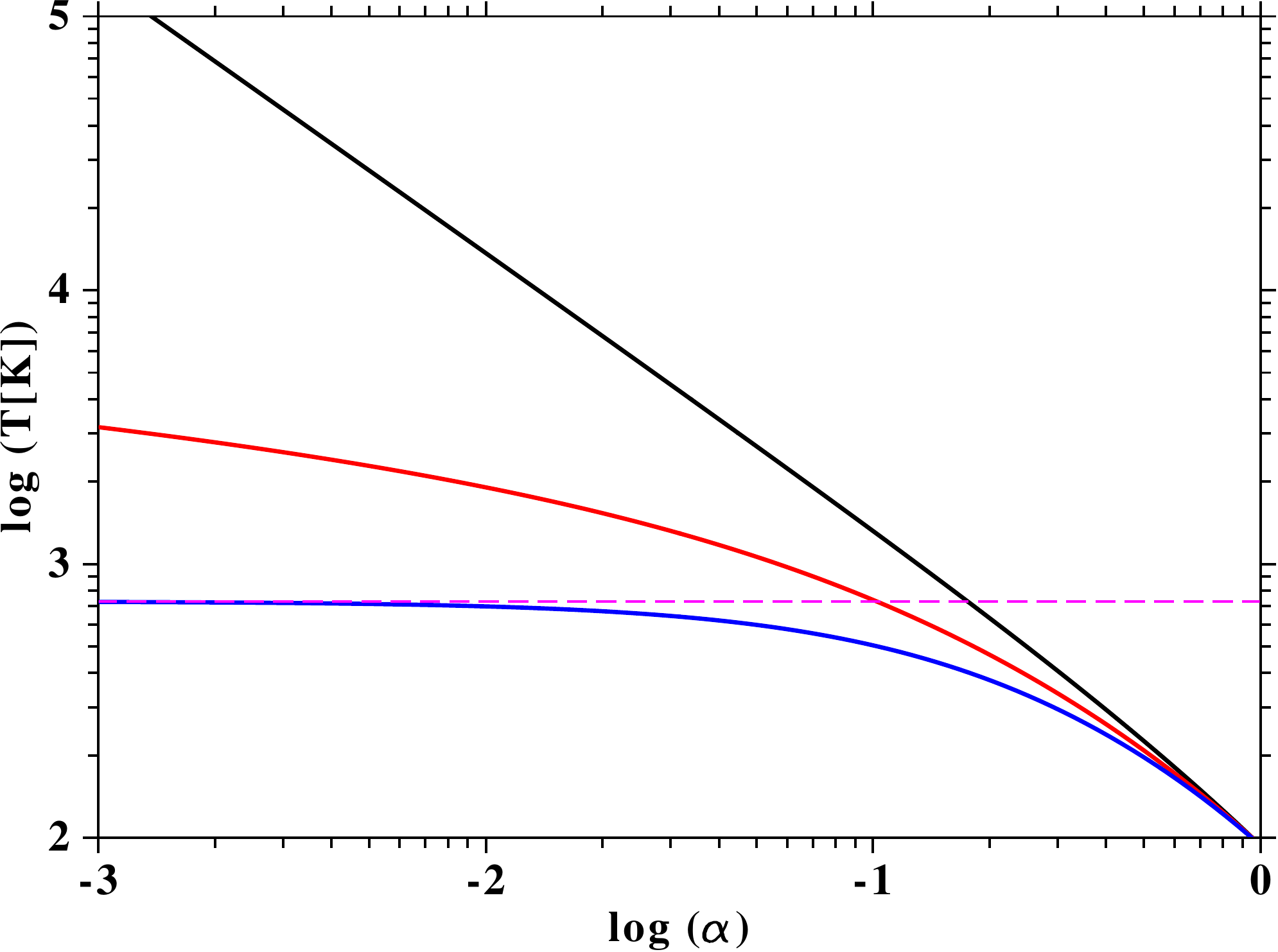}
  \caption{Upper panel: The temperature as a function of $\kappa$ as defined in Eq.~\ref{eq:temp} 
	for $\Theta=1$\,km/s, and $\alpha = 0.1$ (red line) and $\alpha=0.01$ (green line), by comparison 
	with the SKD temperature (black line) with its pole at $\kappa=3/2$ (black dashed vertical line), 
	and the RMD temperature (dashed blue line).
  Lower panel: Variation of the temperature with $\alpha$. The black, red and blue lines correspond to
    $\kappa = 1, 1.5, 2$, respectively. The dashed magenta (horizontal) line is the SKD temperature for 
		$\kappa=2$.  \label{fig:3}}
\end{figure}

The temperature $T_{r\kappa}$ as a function of $\Theta$, $\kappa$ and $\alpha$ 
is more complicated than $T_\kappa$ and $T_M$, and these are contrasted in Fig.~\ref{fig:3}.
The RKD has continuously finite temperature values below $\kappa=3/2$
({\bf upper panel, Fig.~\ref{fig:3}}), while the SKD temperature is 
not defined in that range. 
Moreover, for moderately high values of $\kappa$ (lower panel) like the ones reported by the observations 
in the solar wind \cite{Pierrard-Lazar-2010}, e.g., $2 \leqslant \kappa \leqslant 6$ (far enough from the 
unphysical pole at $\kappa=3/2$) the temperature of RKD and SKD converge. For even higher values of $\kappa \to 
\infty$ the SKD and the RKD approach the Maxwellian limit.
In the lower panel the blue line is the temperature of RKD for $\kappa =2$ and it is almost identical with 
that of SKD if $\alpha$ is sufficiently small, i.e., $\alpha < 0.01$. 
For lower values of $\alpha$ the lines (blue, red, black) diverge when $\kappa$ decreases, while for higher 
values of $\alpha \rightarrow 1$ they converge to the RMD temperature.
Thus, Fig.~\ref{fig:3} illustrates why the RKD is a regularizing generalization of 
the SKD. While the SKD yields diverging temperature for $\kappa \leqslant 3/2$ and, 
thus,  overestimates the temperature for $\kappa$-values approaching this pole, the RKD keeps the temperature 
at finite values for all $\kappa>0$. 
Furthermore, for the same $\kappa$-values indicated by the observations
($\kappa \gg 3/2$) and a sufficiently low parameter $\alpha$, the deviations of the RKD from the SKD are always small.

%%%%%%%%%%%%%%%%%%%%%%%%%%%%%%%%%%%%%%%%%%%%%%%%%%%
\section{Motivation for an exponential cut-off}
%%%%%%%%%%%%%%%%%%%%%%%%%%%%%%%%%%%%%%%%%%%%%%%%%%%

Pure power law distributions in the particle velocity (or the momentum or the energy) are often observed 
in the context of acceleration processes. Such power laws are not well-posed for use over the whole
velocity interval ($[0,\infty]$ in the classical or $[0,c]$ in the relativistic case, with $c$
the speed of light): in the limit $v \to 0$ the distributions diverge and in the limits $v \to \infty$
or $v \to c$ they correspond to diverging moments or unphysical boundary behaviour, respectively. While 
the SKD does not have the first defect, it is hampered by the second. Obviously, the existence of all 
velocity moments requires an exponential cut-off.

In the field of origin of $\kappa$-distributions,
i.e.\ space physics, there are various cases of such exponential
cut-offs of power laws.
{\bf Particularly eloquent are the electron spectra, e.g., Figure~6 in
  \cite{Lin-1998}, with evidences of two or three components (core,
  halo and superhalo) and each of them suggesting an exponential
  cutoff.} We
discuss two examples that also reveal the smallness of the cut-off
parameter $\alpha$.

%%%%%%%%%%%%%%%%%%%%%%%%%%%%%%%%%%%%%%%%%%%%%%%%%%%%%%%%
\subsection{Acceleration in the interplanetary medium}
%%%%%%%%%%%%%%%%%%%%%%%%%%%%%%%%%%%%%%%%%%%%%%%%%%%%%%%%

Fisk \& Gloeckler\cite{Fisk-Gloeckler-2012} have developed a theory based on rarefaction and compression waves 
in the solar wind in order to explain the frequently observed so-called $v^{-5}$-distributions of suprathermal 
particles. Their derivation resulted in distribution functions of the form 
\begin{eqnarray}
  \label{eq:fg1}
  f_{FG} =
  f_{0,th}\left({v}/{v_{th}}\right)^{-5}\exp\left\{-{9D}/{((1+\beta)^{2}\delta u^{2}t})\right\}, \;\;\; 
\end{eqnarray}
i.e.\ power laws with an exponential cut-off. In the above formula $D=D_{0}v^{1+\beta}$ is the spatial 
diffusion coefficient, $v_{th}\approx 30$\,km/s is the thermal speed, $\delta u^2$ quantifies fluctuations 
in velocity, and $t$ is time. This form corresponds to the limit discussed in Eq.~\ref{eq:pl1} above, 
i.e.\ it is an RKD in the limit $v>\kappa\Theta$.

Assuming $\beta=1$ to obtain the standard $v^{2}$ dependence of $D$ and with the velocity diffusion coefficient
$D_{vv}= \delta u^{2} v^{2}/(9 D)$ we find 
\begin{eqnarray}
  \label{eq:fg2}
  \frac{9D}{4 \delta u^{2} t} = \frac{9D v^{2}}{4 \delta u^{2} t
  v^{2}} = \frac{v^{2}}{4 D_{vv}t} = \frac{v^{2}}{4
  v_{typ}^{2}} = \alpha^{2} \frac{v^{2}}{\Theta^{2}},
\end{eqnarray}
where $v_{typ}$ is a typical speed, assumed to be a multiple of $\Theta$ and 
$\alpha = {\Theta^{2}}/{4 v_{typ}^{2}} = (4 m^{2})^{-1}$, with an integer $m$. Hence, 
$\alpha$ is indeed a small parameter less than unity.

%%%%%%%%%%%%%%%%%%%%%%%%%%%%%%%%%%%%%%%%%%%%%%%%%%%%%%%%%%%%%%%
\subsection{Acceleration at the solar wind termination shock}
%%%%%%%%%%%%%%%%%%%%%%%%%%%%%%%%%%%%%%%%%%%%%%%%%%%%%%%%%%%%%%%

Steenberg \& Moraal\cite{Steenberg-Moraal-1999} used a distribution function of the type
\begin{eqnarray}
  \label{eq:sm1}
  f\propto \left({E_{kin}}/{E_{c}}\right)^{-q} 
           \exp\left\{-b  \left({E_{kin}}/{E_{c}}\right)^{2a}\right\}
\end{eqnarray}
to model the anomalous cosmic ray flux at the solar wind termination shock. Here $E_{kin}$ is the
kinetic and $E_{c}$ a cut-off energy, with $q>0$. The authors did fit this expression to data and determined 
the parameters $a,b$ to be of the order of unity and 0.2, respectively. Thus, Eq.(\ref{eq:sm1} 
resembles again an RKD with an $\alpha < 1$.

%%%%%%%%%%%%%%%%%%%%%%%%%%%%%%%%%%%%%%%%%%
\section{On the persistence of kinetic results}
%%%%%%%%%%%%%%%%%%%%%%%%%%%%%%%%%%%%%%%%%%

\begin{figure} 
\hspace*{0.55cm}
\includegraphics[width=0.50\textwidth]{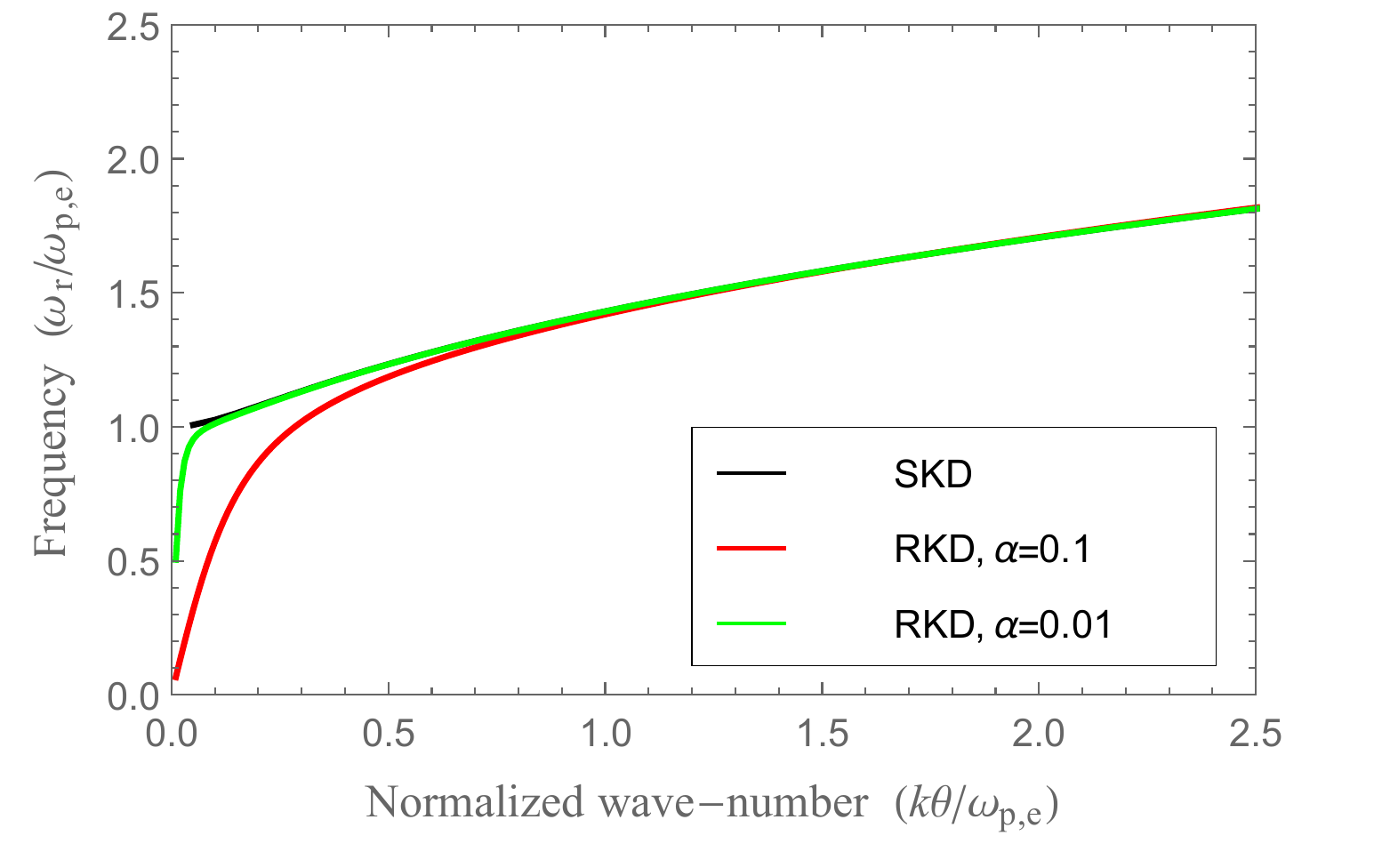}
\includegraphics[width=0.50\textwidth]{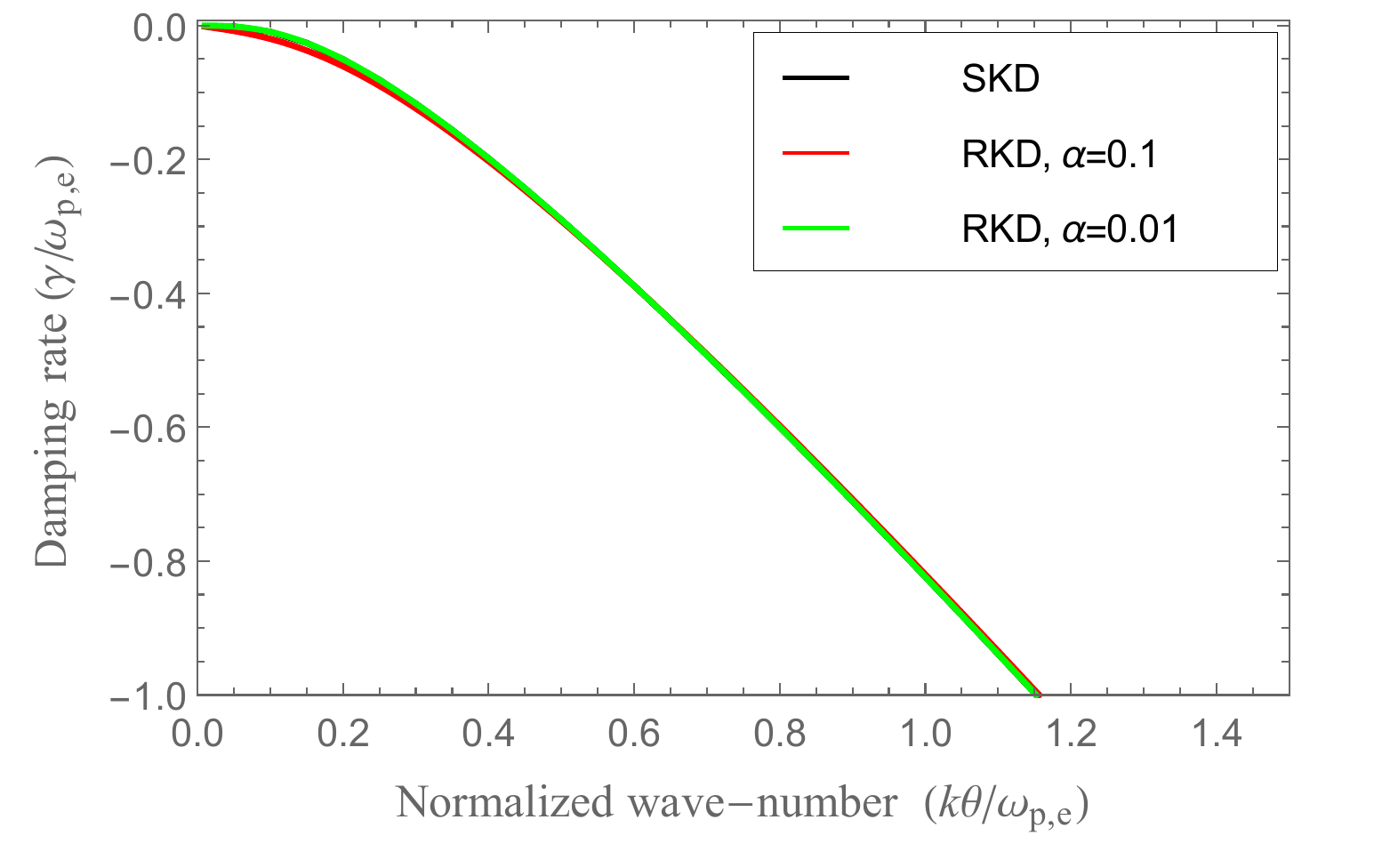}
\caption{Langmuir modes dispersion curves $\omega_r (k)$ (left) and damping rates $\gamma(k)$ 
(right) for the SKD and the RKD.} \label{fig:4}  
\end{figure}

In order to illustrate the influence on the results that are derived within the framework of kinetic 
theory, we consider (electrostatic) Langmuir waves. When using the RKD, these waves have the dispersion relation
\be
0 = 1 - {\omega_{pe}^2 \over k^2 \theta^2} Z'_{\kappa,m} (\xi), \;\; {\rm where} \;\; \xi = {\omega \over k \theta},  
\label{e1}
\ee
and $Z^r_{r,\kappa} (\xi) = \partial Z_{r,\kappa} (\xi) / \partial \xi$ 
is the derivative of what we can call the regularized $\kappa$-dispersion function
\be
Z_{r,\kappa} (\xi) = {1 \over \kappa^{1/2} I_0} {1 \over \pi^{1/2}} \int_{-\infty}^{\infty} {du \over u-\xi} \;
{\exp\{-\alpha^2 u^2\} \over \left(1+ u^2 / \kappa \right)^\kappa},
\ee
with
\begin{eqnarray}
I_0 &=& {\Gamma(\kappa-1/2) \over \Gamma(\kappa)} \, \mbox{$_1$F$_1$}\left({1\over 2}, {3\over 2}- \kappa; 
\kappa \alpha^2 \right)\\\nonumber &&+{\Gamma(1/2 - \kappa) \over \pi^{1/2}} (\kappa
   \alpha^2)^{\kappa-1/2}\, \mbox{$_1$F$_1$} \left(\kappa,\kappa+{1\over 2}; 
\kappa \alpha^2 \right).
\end{eqnarray}
For the derivative the explicit form reads
\be 
Z'_{r,\kappa} (\xi) = - 2 \left\{\alpha^2+ {I \over I_0} + \xi \left[\alpha^2 Z_{r,\kappa} (\xi) + Z_{r,\kappa+1} (\xi) 
\right] \right\}
\ee
where 
\begin{eqnarray}
  I &=& {\Gamma(\kappa+1/2) \over \Gamma(\kappa+1)} \, \mbox{$_1$F$_1$} \left({1\over 2}, {1\over 2}- \kappa; \kappa \alpha^2 \right)\\\nonumber
&&-{\Gamma(1/2 - \kappa) \over \pi^{1/2}(1/2+\kappa)} (\kappa
   \alpha^2)^{\kappa+1/2}\\\nonumber
  &&\hspace*{2cm} _1F_1\left(\kappa+1,\kappa+{3\over 2}; 
\kappa \alpha^2 \right),
\end{eqnarray}
and use was made of the identity $\Gamma(-1/2 - \kappa) = - \Gamma(1/2 - \kappa)/(1/2 + \kappa)$.
Fig.~\ref{fig:4} displays the wave-number dispersion curves of the (normalised) frequencies $\omega_r(k) 
= \Re(\omega)(k)$ and damping rates $\gamma(k) = \Im(\omega)(k)$ for the SKD with $\alpha=0$ and two 
RKDs with $\alpha=0.1$ and $0.01$. 
Evidently, the RKD gives mostly the same values as the SKD, the differences being limited to low wave-numbers 
and decrease with decreasing cut-off parameter $\alpha$.

%%%%%%%%%%%%%%%%%%%%%%%
\section{Conclusion}
%%%%%%%%%%%%%%%%%%%%%%%

We defined a regularized $\kappa$-distribution (RKD) that, by contrast to the 
standard $\kappa$-distribution (SKD), has non-diverging velocity moments and is well-defined for all 
positive $\kappa$. Not only does the RKD remove all singularities from the theory of the SKD but it 
also allows to calculate all velocity moments analytically.
The exponential cut-off of the RKD is motivated from theoretical considerations and from measurements. Both 
allow to determine the cut-off parameter to be significantly lower than unity. With such small values, fits 
to distributions of data are not sensitive to whether an SKD or an RKD is used, 
because the cut-off can be introduced beyond the observational
threshold (of a given instrument).

Perspectives are now opened by a RKD to approach the plasma micro-states for low 
$\kappa$-values (i.e., below the limit $\kappa \leqslant 3/2$ imposed
for a SKD), which have been excluded from the observational reports by reason of unrealistic parameter 
results \cite{Stverak-etal-2008} so far, although there is recently
some indication of low-$\kappa$ power laws\cite{Desai-etal-2016}.
For higher $\kappa > 3/2$ kinetic effects remain easily accessible by a SKD, while a macroscopic 
(fluid-like) description requires switching to the RKD.
The moments of the SKD and the RKD differ significantly, so that any result obtained within a 
fluid approach must be expected to depend on this choice \cite{Fahr-etal-2014}. In practice, only the RKD 
allows a well-defined fluid approach, because only the RKD admits a complete (infinite) set of velocity 
moments that are well-defined for all positive $\kappa$-values.

%%%%%%%%%%%%%%%%%%%%%%%
\section*{Appendix}
%%%%%%%%%%%%%%%%%%%%%%%
{\bf For convenience the two used generalized hypergeometric series
  (or functions)
  $_{1}F_{0}([a],[];x)$ and $_{1}F_{1}([],[b];x)$ are
  given here
  \begin{eqnarray}\nonumber
    _{1}F_{0}([a],[];z)&=& \sum\limits_{n=0}^{\infty}(a)_{n}\,\frac{z^{n}}{n!} \\\nonumber
    _{1}F_{1}([a],[b];z) &=& \sum\limits_{n=0}^{\infty}\frac{(a)_{n}}{(b)_{n}}\,\frac{z^{n}}{n!}
  \end{eqnarray}\nonumber
  with $a,b\in\mathbb{R} \mathrm{\ and\ }z\in\mathbb{C}$,
  where $b\notin \{0,-1,...,-m\},$ $ \ m\in \mathbb{N}$ and
  $(\bullet)_{n}$ is the Pochhammer symbol defined as
  \begin{eqnarray}\nonumber
    (a)_{0} = 1 \\\nonumber
    (a)_{n} = a(a+1)(a+2)...(a+n-1) \qquad n>0, n\in\mathbb{N}
  \end{eqnarray}
  Another notation for the confluent hypergeometric function $_{1}F_{1}([a],[b];z)$ is the Kummer function
  $M(a,b;x)$.
  For more details see  \cite{Abramowitz-Stegun-1972} or \cite{Gradstein-Ryshik-1981}.

}
\ack
KS and HF are grateful to the
  \emph{Deut\-sche For\-schungs\-ge\-mein\-schaft} (DFG) funding the
  projects FI706/15-1 and SCHE334/9-2. ML acknowledges funding from
  the DFG - project SCHL~201/35-1, and FWO--Vlaanderen - project G0A2316N.
This paper
results from a collaboration of the authors initiated by the ISSI project: ``Kappa
Distributions: From Observational Evidences via Controversial Predictions to a Consistent Theory of
Suprathermal Space Plasmas.''

\vspace{1cm}
%\bibliographystyle{iopart-num}
%\bibliography{references}

\begin{thebibliography}{10}
\expandafter\ifx\csname url\endcsname\relax
  \def\url#1{{\tt #1}}\fi
\expandafter\ifx\csname urlprefix\endcsname\relax\def\urlprefix{URL }\fi
\providecommand{\eprint}[2][]{\url{#2}}

\bibitem{Olbert-1968}
{Olbert} S 1968 {Summary of Experimental Results from M.I.T. Detector on IMP-1}
  {\em Physics of the Magnetosphere\/} ({\em Astrophysics and Space Science
  Library\/} vol~10) ed {Carovillano} R~D~L and {McClay} J~F p 641

\bibitem{Vasyliunas-1968}
{Vasyliunas} V~M 1968 {\em J. Geophys. Res.\/} {\bf 73} 2839--2884

\bibitem{Treumann-Jaroschek-2008}
{Treumann} R~A and {Jaroschek} C~H 2008 {\em Physical Review Letters\/} {\bf
  100} 155005 (\textit{Preprint} \eprint{0711.1676})

\bibitem{Pierrard-Lazar-2010}
{Pierrard} V and {Lazar} M 2010 {\em Sol.\ Phys.\/} {\bf 267} 153--174
  (\textit{Preprint} \eprint{1003.3532})

\bibitem{Beck2017}
Beck C and Cohen E 2017 Chapter 6: Superstatistics: Superposition of
  maxwell–-boltzmann distributions {\em Kappa Distributions\/} ed Livadiotis
  G (Elsevier) pp 313 -- 328 ISBN 978-0-12-804638-8
  \urlprefix\url{https://www.sciencedirect.com/science/article/pii/B9780128046388000061}

\bibitem{Webb-etal-2012}
{Webb} S, {Litvinenko} V~N and {Wang} G 2012 {\em Phys. Rev. STAB\/} {\bf 15}
  080701 (\textit{Preprint} \eprint{1105.0412})

\bibitem{Elkamash-Kourakis-2016}
{Elkamash} I~S and {Kourakis} I 2016 {\em Phys. Rev. E\/} {\bf 94} 053202

\bibitem{Gosset-1908}
{{Student = Gosset}, WS} 1908 {\em Biometrika\/} {\bf 6} 1--25 ISSN 0006-3444
  (print), 1464-3510 (electronic)
  \urlprefix\url{http://www.jstor.org/stable/2331554}

\bibitem{Pearson-1916}
{{Pearson}, K} 1916 {\em Philosophical Transactions of the Royal Society of
  London A: Mathematical, Physical and Engineering Sciences\/} {\bf 216}
  429--457 ISSN 0264-3952 (\textit{Preprint}
  \eprint{http://rsta.royalsocietypublishing.org/content/216/538-548/429.full.pdf})
  \urlprefix\url{http://rsta.royalsocietypublishing.org/content/216/538-548/429}

\bibitem{Abdul-Mace-2014}
{Abdul} R~F and {Mace} R~L 2014 {\em Comp. Phys. Comm.\/} {\bf 185} 2383--2386

\bibitem{Lazar-etal-2016}
{Lazar} M, {Fichtner} H and {Yoon} P~H 2016 {\em Astron. Astrophys.\/} {\bf
  589} A39 (\textit{Preprint} \eprint{1602.04132})

\bibitem{Ziebell-Gaelzer-2017}
{Ziebell} L~F and {Gaelzer} R 2017 {\em Phys. Plas.\/} {\bf 24} 102108

\bibitem{Lazar-etal-2017}
{Lazar} M, {Pierrard} V, {Shaaban} S~M, {Fichtner} H and {Poedts} S 2017 {\em
  Astron. Astrophys.\/} {\bf 602} A44 (\textit{Preprint} \eprint{1703.01459})

\bibitem{Lazar-etal-2015}
{Lazar} M, {Poedts} S and {Fichtner} H 2015 {\em AA\/} {\bf 582} A124

\bibitem{Leubner-2000}
{Leubner} M~P 2000 {\em Planet. Space Sci.\/} {\bf 48} 133--141

\bibitem{Langmayr-2005}
{Langmayr} D, {Biernat} H~K and {Erkaev} N~V 2005 {\em Physics of Plasmas\/}
  {\bf 12} 102103

\bibitem{Yoon-2014}
{Yoon} P~H 2014 {\em Journal of Geophysical Research (Space Physics)\/} {\bf
  119} 7074--7087

\bibitem{Lin-1998}
{Lin} R~P 1998 {\em Space Sci. Rev.\/} {\bf 86} 61--78

\bibitem{Fisk-Gloeckler-2012}
{Fisk} L~A and {Gloeckler} G 2012 {\em Space Sci. Rev.\/} {\bf 173} 433--458

\bibitem{Steenberg-Moraal-1999}
{Steenberg} C~D and {Moraal} H 1999 {\em J. Geophys. Res.\/} {\bf 104}
  24879--24884

\bibitem{Stverak-etal-2008}
{{\v S}tver{\'a}K} {\v S}, {Tr{\'a}vn{\'{\i}}{\v c}ek} P, {Maksimovic} M,
  {Marsch} E, {Fazakerley} A~N and {Scime} E~E 2008 {\em Journal of Geophysical
  Research (Space Physics)\/} {\bf 113} A03103

\bibitem{Desai-etal-2016}
{Desai} M~I, {Mason} G~M, {Dayeh} M~A, {Ebert} R~W, {McComas} D~J, {Li} G,
  {Cohen} C~M~S, {Mewaldt} R~A, {Schwadron} N~A and {Smith} C~W 2016 {\em
  Astrophys. J.\/} {\bf 828} 106 (\textit{Preprint} \eprint{1605.03922})

\bibitem{Fahr-etal-2014}
{Fahr} H~J, {Fichtner} H and {Scherer} K 2014 {\em J. Geophys. Res.\/} {\bf
  119} 7998--8005

\bibitem{Abramowitz-Stegun-1972}
{Abramowitz} M and {Stegun} I~A 1972 {\em {Handbook of Mathematical
  Functions}\/} (New York: Dover, 1972)

\bibitem{Gradstein-Ryshik-1981}
{Gradshteyn} I~S and {Ryzhik} I~M 2007 {\em Table of integrals, series, and
  products\/} (Elsevier/Academic Press, Amsterdam) ISBN 978-0-12-373637-6;
  0-12-373637-4

\end{thebibliography}
\providecommand{\newblock}{}

\end{document}